\documentclass[13pt]{iopart}

\usepackage{iopams}
\expandafter\let\csname equation*\endcsname\relax
\expandafter\let\csname endequation*\endcsname\relax
\usepackage{amsmath}
\usepackage{graphicx}
\usepackage{xcolor}
\usepackage{amsmath, amsthm, amssymb, amsfonts}
\usepackage{amstext}
\usepackage{amsmath,bm}
\usepackage{url}
\usepackage{color} 
\usepackage{adjustbox}
\usepackage{float}
\usepackage{lipsum}
\begin{document}

\title[Re-examination of $\beta$-decay in Hg, Pb and Po Isotopes]{Re-examination of $\beta$-decay in Hg, Pb and Po Isotopes}
\author{Jameel-Un Nabi$^{1,2}$, Tuncay Bayram$^{3}$, Muhammad Riaz$^{4,*}$, Asim Ullah$^{2}$, Anes Hayder$^{3}$, Şevki Şenturk$^{3}$, and Mahmut B\"{o}y\"{u}kata$^{5}$}

\address{$^{1}$University of Wah, Quaid Avenue, Wah Cantt 47040, Punjab, Pakistan.}
\address{$^{2}$Faculty of Engineering Sciences, Ghulam Ishaq Khan Institute of Engineering Sciences and Technology, Topi, 23640, KP, Pakistan.}
\address{$^{3}$Department of Physics, Faculty of Science, Karadeniz Technical University, Trabzon, 61080, Türkiye.}
\address{$^{4}$Department of Physics, University of Jhang, Jhang 35200, Punjab, Pakistan.}
\address{$^{5}$Physics Department, Science and Arts Faculty, K\i r\i kkale
	University, 71450, K\i r\i kkale, Türkiye.}
\address{$^*$ Corresponding author}
\ead{ jameel@uow.edu.pk, t.bayram@ymail.com, mriaz@uoj.edu.pk, asimullah844@gmail.com, anes.m.s.q@gmail.com, sevkisenturk61@gmail.com and boyukata@kku.edu.tr}
\vspace{10pt}
\begin{abstract}
	This study re-examines the effect of nuclear deformation on the \textbf{calculated} Gamow-Teller (GT)  strength distributions of neutron-deficient  ($^{178-192}$Hg, $^{185-194}$Pb and $^{196-206}$Po) nuclei. The nuclear ground state properties and shape parameters were calculated using the Relativistic Mean Field  model. Three  different density-dependent interactions were used in the calculation. Estimated shape parameters were later used within the framework of deformed proton-neutron quasi-random phase approximations  model, with a separable interaction, to calculate the  GT strength distributions, half-lives and branching ratios for these neutron--deficient isotopes. It was concluded that half-lives and GT strength distributions vary considerably with change in shape parameter.

\end{abstract}
%
\vspace{2pc}
\ioptwocol
\maketitle
\section{Introduction}
\label{intro}
The study of nuclear structure properties of neutron-deficient nuclei is  one of the most active areas of research in nuclear physics. The researchers, on one side, endeavor to upgrade the  experimental facilities, while on the other side, develop new theoretical tools to extend the knowledge of nuclear physics far away from the $\beta$-stability line. The existence of different shapes of a particular specie of nuclei is a proven phenomenon. Nuclei change their shapes gradually from spherical to deformed by the polarizing effect of added nucleons \cite{Pan21}. The coexistence of varying collective nuclear shapes may give rise to several 0$^+$ states at low energies. Nuclear properties change as a nucleus exhibit varying shapes. Such studies are indeed useful for a better understanding of the nucleus.  

The shape coexistence of nuclei at low energy plays a crucial role in characterizing the neutron-deficient Hg, Pb, and Po isotopes \cite{Jul01}. Shape coexistence in mid-shell nuclei, having Z $=$ 50 to 82 shells, requires careful investigations. In some nuclei, spherical and deformed intruder states coexist closely \cite{Woo92}.
The first shape coexistence in neutron-deficient nuclei was observed during the isotope shift measurements for Hg isotopes. {Transitions in the ground states of the chain of odd-A mercury isotopes and from an oblate to a prolate shape in the ground state band of even mass Hg isotopes were investigated \cite{Fra75,Bon76}}. Marsh et. al.~\cite{Mar18} studied shape staggering effects by combining the experimental data with theoretical calculations. Neutron-deficient $^{177-185}$Hg isotopes were studied using in-source laser resonance-ionization spectroscopy. The results were  compared with calculations using the density functional theory and Monte Carlo shell model. The conclusions supported deformation effects in mercury isotopes \cite{Sel19}. The configuration mixing of Hg nuclei was studied by using  $\gamma-\gamma$ electronic fast timing techniques with the { LaBr$_3$(Ce) detector array of the GRIFFIN spectrometer performed at TRIUMF-ISAC facility \cite{Ola19} and at the Tandem Van de Graaff accelerator in Germany} \cite{Esm18}. 

$^{186}$Pb is the only identified nucleus where two first excited states within 700 keV of the ground state form a triplet of zero-spin states~\cite{And00}. These are assigned
	with prolate, oblate and spherical shapes. Ojala and collaborators ~\cite{Joo22} reported on a precision measurement of
	the properties of collective transitions in $^{186}$Pb. The authors studied the nucleus $^{186}$Pb in a simultaneous  in-beam $\gamma$-ray and electron spectroscopy experiment employing the recoil-decay tagging technique for their measurements. The findings of this study suggested that the
	0$_2^+$ state has a large prolate admixture. The study further allowed for firm
	assignments of the 2$_2^+$, 4$_2^+$, 6$_2^+$ and 8$_2^+$ states that form part of the
	predominantly oblate band. 
	
The deformation and mixing of coexisting shapes in neutron-deficient Po isotopes were investigated  using Coulomb excitation experiments with post-accelerated beams of neutron-deficient $^{196,198,200,202}$Po isotopes at the REX-ISOLDE facility~\cite{Kes15}. The experimental results confirmed the onset of deformation from $^{196}$Po onward.  The investigation hinted at the mixing of a spherical structure with a weakly deformed rotational structure.


Shape coexistence was predicted in Hg and Pb isotopes by phenomenological mean field models and the Strutinsky method \cite{Ben89}. Later these findings were validated by more sophisticated calculations  \cite{Smi03,Nik02,Lib99,Egi04,Ben04}. Earlier the $\beta$-decay properties
of the neutron-deficient Hg, Pb and Po isotopes were studied  using a deformed
Skyrme HF+BCS+QRPA approach~\cite{Mor06}. The authors 
commented that an accurate calculation
of the excitation energies of the 0$^+$ states was beyond the
scope of their theoretical framework.  Only estimated values of deformation parameter ($\beta_2$) were incorporated in their study for even-even cases. They estimated oblate and prolate minima for Hg and Po isotopes. For  Pb nuclei, oblate, prolate and spherical shapes were deduced (except $^{194}$Pb where only spherical and oblate shapes were reported). 
The authors found out that only the GT strength distributions altered significantly with the $\beta_2$ values.  The study concluded 
that neither the half-lives nor the summed strengths were good
observables to study the deformation effects. The current study re-examines the findings of Ref.~\cite{Mor06}. Different sets of deformation parameter were used in our investigation. {The $\beta_2$ value was incorporated as a free parameter in our nuclear model to calculate GT strength distributions and $\beta$-decay half-lives of neutron-deficient Hg, Pb and Po isotopes.}

We first calculated the ground state properties of Hg, Pb and Po isotopes. The  Relativistic Mean
Field (RMF) model, using three different versions of
density-dependent RMF functionals, was employed to calculate these properties. Details of the functionals would be provided in the next section.  Quadrupole moment constrained RMF calculation was performed to
obtain potential energy curves (PECs) as a function of the  ground state quadrupole deformation parameter. Minima of the curves provided us a set of deformation parameters for these neutron-deficient isotopes. 
A second set of $\beta_2$ parameters  using the Finite Range Droplet Model (FRDM)~\cite{moller16} was incorporated in our investigation.  For even-even isotopes, deformation parameters using the Skyrme HF+BCS+QRPA~\cite{Mor06} were also employed in this project. 
The proton-neutron quasiparticle random phase
approximation (pn-QRPA) calculation, possessing a simple pairing plus quadrupole Hamiltonian with a schematic
and separable spin-isospin residual interaction, was later performed for all sets of deformation parameters to re-examine their effect on calculated GT strength distributions and $\beta$-decay half-lives.  
 
\section{Formalism}
\label{sec:1}
Nuclear structure calculations were performed using the Relativistic Mean Field (RMF) model. We calculated the ground state binding energies per nucleon and potential energy curves (PECs). Three different versions of
density-dependent RMF functional were used in our calculation. The calculations of GT strength distributions and half-lives were performed using the pn-QRPA model. {The necessary formalism of the two models,  used in our investigation,} is described in the succeeding sub-sections.

\subsection{The Relativistic Mean Field (RMF) Model}\label{subsection 1}

Phenomenologically, the nucleons interact with each other via exchange of various mesons and photons in the RMF model~\cite{walecka1974,boguta1977,ring1996,lalazissis1999,vretenar2005,lalazissis2005,Bay13,bayram19}. 

In this investigation, density-dependent point-coupling and meson-exchange types of RMF model have been used for the calculation of ground state properties of nuclei. In this subsection, we will briefly describe the density-dependent meson-exchange  and point-coupling versions of the RMF model. Details about other forms of the RMF model could be seen in Ref.~\cite{meng2006}. 
 
The starting point of density-dependent meson-exchange RMF model is a Lagrangian density

\begin{equation}
	\mathcal{L}=\mathcal{L}_N+\mathcal{L}_m+\mathcal{L}_{int},\label{lagden}
\end{equation}  
where $\mathcal{L}_N$ is the Lagrangian for the free nucleon given by

\begin{equation}
	\mathcal{L}_N=\bar{\Psi}(i\gamma_\mu\partial^\mu-m)\Psi.\label{lagnuc}
\end{equation}
In Eq. (\ref{lagnuc}), $\psi$ represents the Dirac spinor and $m$ is the nucleon mass. $\mathcal{L}_m$ term in Eq.~(\ref{lagden}) is the Lagrangian for free meson  and electromagnetic fields

\begin{equation}
	\begin{split}
		\mathcal{L}_m &=\frac{1}{2}\partial_\mu\sigma\partial^\mu\sigma-\frac{1}{2}m_\sigma^2\sigma^2-\frac{1}{2}\Omega_{\mu\nu}\Omega^{\mu\nu}+\frac{1}{2}m^2_\omega\omega_\mu\omega^\mu \\ 
		&-\frac{1}{4}\overrightarrow{R}_{\mu\nu}.\overrightarrow{R}^{\mu\nu}+\frac{1}{2}m_\rho^2 \overrightarrow{\rho}_\mu.\overrightarrow{\rho}^\mu -\frac{1}{4}F_{\mu\nu}F^{\mu\nu}, \label{lagmes}
	\end{split}
\end{equation}  
where arrows represent isovectors. The masses of $\sigma$, $\omega$ and $\rho$ mesons are denoted by $m_\sigma$, $m_\omega$ and $m_\rho$, respectively. Field tensors are given by the following equations 

\begin{equation}
	\begin{split}
		\Omega_{\mu\nu}=\partial_\mu\omega_\nu-\partial_\nu\omega_\mu, \\
		\overrightarrow{R}_{\mu\nu}=\partial^\mu\overrightarrow{\rho}_\nu-\partial_\nu\overrightarrow{\rho}_\mu, \\
		F_{\mu\nu}=\partial_\mu A_\nu-\partial_\nu A_\mu  \label{fieldtensors}.
	\end{split}
\end{equation}
$\mathcal{L}_{int}$ term in Eq.~(\ref{lagden}) includes meson-nucleon and photon-nucleon interactions which can be expressed as 

\begin{equation}
	\begin{split}
		\mathcal{L}_{int}&= -g_\sigma\bar{\Psi}\Psi \sigma - g_\omega\bar{\Psi}\gamma^\mu\Psi\omega_\mu - g_\rho\bar{\Psi}\overrightarrow{\tau}\gamma^\mu\Psi.\overrightarrow{\rho}_\mu-e\bar{\Psi}\gamma^\mu\Psi A_\mu,
	\end{split}
	\label{lagrangian}
\end{equation}
where  $g_\sigma$, $g_\omega$ and $g_\rho$ are the coupling constants of the related mesons.

For the static case, the Hamiltonian density is given by~\cite{ring1996}

\begin{equation}
	\begin{split}
		\mathcal{H}({\bf r}) &=\Sigma_i^\dagger(\bm{\alpha p}+\beta m)\Psi_i \\
		&+ \frac{1}{2}\left[ ({\bf \nabla}\sigma)^2 + m_\sigma^2\sigma^2\right] - \frac{1}{2}\left[ ({\bf \nabla}\omega)^2 + m_\omega^2\omega^2\right] \\ 
		&-\frac{1}{2}\left[ ({\bf \nabla}\rho)^2 + m_\rho^2\rho^2\right] - \frac{1}{2}({\nabla \bf A})^2 \\
		&+ \left[ g_\sigma\rho_s\sigma+g_\omega j_\mu\omega^\mu + g_\rho \overrightarrow{j}_{\mu}.\overrightarrow{\rho}^{\mu} + ej_{p\mu}A^{\mu}  \right], \label{hamden}
	\end{split}
\end{equation}  
where density and currents are given by  

\begin{equation}
	\rho_s({\bf r})=\Sigma_{i=1}^A \bar{\Psi}_i({\bf r}) \Psi_i({\bf r}) \label{isoscalar}
\end{equation}

\begin{equation}
	j_\mu({\bf r})=\Sigma_{i=1}^A \bar{\Psi}_i({\bf r}) \gamma_\mu \Psi_i({\bf r}) \label{isovector}
\end{equation}

\begin{equation}
	\overrightarrow{j}_\mu({\bf r})=\Sigma_{i=1}^A \bar{\Psi}_i({\bf r}) \overrightarrow{\tau}\gamma_\mu \Psi_i({\bf r}) \label{isovectorvec}
\end{equation}

\begin{equation}
	j_{p\mu}({\bf r})=\Sigma_{i=1}^Z \bar{\Psi}_i^\dagger({\bf r})\gamma_\mu \Psi_i({\bf r}). \label{electromagnetic}
\end{equation}
The {\it no sea approximation}  can be used for summation in these densities. The total energy can be determined by integrating the Hamiltonian density 

\begin{equation}
	E_{RMF}\left[\Psi,\bar{\Psi},\sigma,\omega^{\mu},\overrightarrow{\rho}^{\mu}, A^{\mu} \right]=\int d^3r \mathcal{H}({\bf r}). \label{toten}
\end{equation}     

The single-nucleon Dirac equation can be obtained by the variation of energy density functional given in Eq.~(\ref{toten}) 

\begin{equation}
	\widehat{h}_D \Psi_i=\epsilon_i\Psi_i. \label{diraceq}
\end{equation} 
In Eq.~(\ref{diraceq}),  $\widehat{h}_D$ is the Dirac Hamiltonian given by

\begin{equation}
	\widehat{h}_D=\bm{\alpha}(\bm{p-\Sigma}) + \Sigma_0 + \beta(m+\Sigma_s), \label{dirham}
\end{equation} 
where $\Sigma$ denotes the single-nucleon self-energies determined by the following equations 
\begin{equation}
	\begin{split}
		\Sigma_s({\bf r}) &=g_\sigma\sigma ({\bf r}), \\ 
		\Sigma_\mu ({\bf r}) &=g_\omega\omega_\mu ({\bf r}) + g_\rho\overrightarrow{\tau}.\overrightarrow{\rho}_\mu({\bf r}) + eA_\mu({\bf r}) + \Sigma_\mu^R({\bf r}). \label{selfener}
	\end{split}
\end{equation}
The rearrangement contribution to the vector self-energy arises because of the density dependence of the vertex functions $g_\sigma$, $g_\omega$, and $g_\rho$.

\begin{equation}
	\Sigma_\mu^R=\frac{j_\mu}{\rho_v}\left( \frac{\partial g_\sigma}{\partial \rho_v}\rho_s\sigma + \frac{\partial g_\omega}{\partial \rho_v} j_v\omega^v + \frac{\partial g_\rho}{\partial \rho_v}\overrightarrow{j}_v.\overrightarrow{\rho}^v \right). \label{rear}
\end{equation}    

The variation of energy density functional, given in Eq.~(\ref{toten}), with respect to the meson fields gives the following Helmholtz equations 

\begin{equation}
	\left[ -\Delta + m_\sigma^2 \right]\sigma=-g_\sigma\rho_s , \label{helm1}
\end{equation} 

\begin{equation}
	\left[ -\Delta + m_\omega^2 \right]\omega^\mu=g_\omega j^\mu , \label{helm2}
\end{equation}

\begin{equation}
	\left[ -\Delta + m_\rho^2 \right]\overrightarrow{\rho}^\mu=g_\rho \overrightarrow{j}^m , \label{helm3}
\end{equation}
and the Poisson equation for the electromagnetic field

\begin{equation}
	-\Delta A^\mu = ej_p^\mu. \label{helm4}
\end{equation} 

There are no currents and the corresponding spatial components of the meson field vanish for  solution of an even-even nucleus. The Dirac equation converts its form given by

\begin{equation}
	\left\lbrace -i{\bm \alpha\nabla}+\beta M^*({\bf r}) + V({\bf r})\right\rbrace \Psi_i({\bf r})=\epsilon_i \Psi_i({\bf r}) , \label{mass1}
\end{equation}
where $M^*({\bf r})=m+g_\sigma \sigma$ is the effective mass and $V({\bf r})$ is the vector potential given by 

\begin{equation}
	V({\bf r})= g_\omega \omega + g_\rho \tau_3 \rho +eA_0 + \Sigma_0^R. \label{vecpot}
\end{equation}
The rearrangement contribution in Eq.~(\ref{rear}) is consequently reduced to 

\begin{equation}
	\Sigma_0^R=\frac{\partial g_\sigma}{\partial \rho_v}\rho_s\sigma + \frac{\partial g_\omega}{\partial \rho_v} \rho_v\omega + \frac{\partial g_\rho}{\partial \rho_v}\rho_{tv}\rho, \label{rear2}
\end{equation} 
where $\rho_{tv}$ represents the isovector density. The meson-nucleon couplings can be parameterized~\cite{typel99,nik02b,typel05}. The couplings of $\sigma$ and $\omega$ mesons to nucleon field reads
\begin{equation}
	g_i(\rho)=g_i(\rho_{sat})f_i(x)~~~\text{for}~i=\sigma,\omega , \label{coup}
\end{equation}
where $\rho_{sat}$ is the baryon density at saturation in symmetric nuclear matter and 
\begin{equation}
	f_i(x)=a_i\frac{1+b_i(x+d_i)^2}{1+c_i(x+d_i)^2}, \label{coups}
\end{equation}
is a function of $x=\rho/\rho_{sat}$. Following constraints are applicable on the non-independent parameters in Eq.~(\ref{coups}): $f_i(1) = 1$, $f_\sigma(1)^{''}=f_\omega^{''}(1)$ and $f_i^{''}(0)=0$. The density dependence of the $\rho$ meson, in exponential form, is given by
\begin{equation}
	g_\rho(\rho)=g_\rho(\rho_{sat})e^{-a_\rho(x-1)}, \label{coup}
\end{equation}
where $\rho(\rho_{sat})$ and $a_\rho$ provide parameterization of the isovector channel. Parameters of meson-nucleon couplings were tuned for fixing their vertex functions in order to calculate the properties of nuclei. 


We used the DIRHB code~\cite{niksic2014}, employing  three different versions of density-dependent functionals, for solving the RMF equations. These include the meson-exchange D$^3$C~\cite{typel05},  DD-ME2~\cite{lalazissis2005}, and point-coupling PC1~\cite{niksic2008} functionals.


In this study, we consider axially symmetric cases of $^{178-192}$Hg, $^{185-194}$Pb and $^{196-206}$Po nuclei for the RMF calculation. For the determination of ground state  deformation parameter ($\beta_2$), quadrupole moment constrained RMF calculation was performed to obtain the PECs as a function of $\beta_2$.  The binding energy, at a desired deformation, was calculated by constraining the expectation value of quadrupole moment $\langle Q_2\rangle$ to a given value $\mu_2$ in the expectation value of the Hamiltonian
\begin{equation}
	\langle H' \rangle=\langle H\rangle + C_\mu (\langle Q_2\rangle-\mu_2)^2, \label{constr} 
\end{equation}
where $C_\mu$ is the constraint multiplier and
\begin{equation}
	\langle Q_2 \rangle =\frac{3}{\sqrt{5\pi}}Ar^2\beta_2, \label{quad3}
\end{equation}
where $A$ is the mass number and $r = R_0A^{1/3}$ ($R_0 = 1.2$ fm). { Pairing correlations are important} for open-shell nuclei. In this investigation, the Bardeen–Cooper–Schrieffer (BCS) approximation was used to handle the pairing correlations.  For the PEC calculation, we used constant G approximation recipe as suggested in Ref.~\cite{Karatzikos}. 


\subsection{The Proton-neutron Quasiparticle Random Phase Approximation (pn-QRPA) Model}\label{ubsection 2}
The Hamiltonian of the pn-QRPA model employed for the calculation of GT strength distributions is given by
\begin{equation} \label{H}
	H^{pnQRPA} = H^{sp} + V^{ph}_{GT} + V^{pp}_{GT} + V^{pair},
\end{equation}
where $H^{sp}$, $V_{GT}^{ph}$, $V_{GT}^{pp}$ and  $V^{pair}$ represent the single-particle Hamiltonian,  particle-hole, particle-particle GT
forces and pairing
force, respectively. The last term of Eq.~\ref{H} was computed using the BCS approximation. The Nilsson model \cite{Nil55} (with incorporation of $\beta_2$) was used for the calculation of single-particle energies and wavefunctions. The oscillator constant for nucleons was computed using $\hbar\omega=\left(45A^{-1/3}-25A^{-2/3}\right)$ MeV (same for protons and neutrons). The Nilsson-potential
parameters were adopted from Ref.~\cite{ragnarson1984}. 
$Q$-values were taken from the recent mass compilation of Ref. \cite{Aud21}. The inclusion of the GT residual force altered the calculated GT strength distributions significantly. The interaction strengths, particle-particle ($\kappa$) and particle-hole ($\chi$), were parameterized in accordance with  1/A$^{0.7}$ dependence as suggested in Ref.~\cite{Hom96}. The position of GT giant resonances (GTGR) is mainly affected by the particle-hole  force. Consequently, the strength of $\chi$  was constrained by the observed GTGR positions.  We selected some isotopes as representative nuclei, where  experimental GT information was available~\cite{Alg21,Agu15}, to adjust the particle-hole strength parameter. The optimized GT interaction parameters resulted in a reasonable comparison of the calculated and observed GT data for the case of  $^{186}$Hg, $^{190}$Pb and $^{192}$Pb (see Fig.~\ref{Rb}). 
The $\beta$-decay partial half-lives were calculated using the equation
\begin{eqnarray}
\resizebox{.9\hsize}{!}{$t_{(1/2)} = \frac{D}{(g_V/g_A)^{-2}f_A(Z, A, E)B_{GT}(E_f)+f_V(Z, A, E)B_F(E_f)},$}
\end{eqnarray}
where $E$ = ($Q$ - $E_f$), value of constant $g_A/g_V$ was taken as -1.254 and D (= ${2\pi^3 \hbar^7 ln2}/{g^2_V m^5_ec^4}$) was taken as 6295 $s$. The phase space integrals for axial vector and vector transitions were represented by $f_A(E, Z, A)$ and $f_V(E, Z, A)$, respectively. $B_{F}$ ($B_{GT}$) denotes the reduced transition probabilities for the Fermi (GT) transitions.
The total $\beta$-decay half-lives were calculated using the equation
\begin{equation}
	T_{1/2} = \left(\sum_{0 \le E_f \le Q} \frac{1}{t_{(1/2)}}\right)^{-1}.
\end{equation}
All transition probabilities to daughter states within the $Q$ window were included in the summation.
Further details on the formalism used in the current calculation may be seen from Ref.~\cite{Nab23}.

\section{Results and Discussion}
We first discuss the nuclear structure calculations and determination of $\beta_2$ for $^{178-192}$Hg, $^{185-194}$Pb and $^{196-206}$Po. To do the needful, RMF model was employed using three different interactions (D$^3$C, DD-ME2 and DD-PC1). Later we investigate the effect of $\beta_2$ on the  calculated GT strength distributions and $\beta$-decay half-lives using the pn-QRPA model. 

Fig.~\ref{bea_all} shows the calculated binding energies per nucleon (BE/A) for Hg, Pb and Po isotopes using the RMF model with density-dependent D$^3$C, DD-PC1 and DD-ME2 interactions. For comparison, the results of the RMF model with non-linear NL3* interaction~\cite{bayram13a}, HFB theory with SLy4 parameter set~\cite{sto03}, FRDM~\cite{moller16} and available experimental data~\cite{Aud21} are also displayed. A maximum deviation of around 0.15 MeV from experimental data was obtained using the D$^3$C functional for isotopic chains of Hg, Pb and Po. The results of FRDM and RMF model with DD-ME2 and DD-PC1 functionals are in good agreement with the experimental data as against those using the HFB method with SLy4 and RMF model with non-linear NL3* interaction.

One of the most important properties of nuclei is the shape parameter. We focused on predicting the ground-state shape  of $^{178-192}$Hg, $^{185-194}$Pb and $^{196-206}$Po. We employed the quadrupole moment constrained axially symmetric RMF+BCS calculation for obtaining the PECs of the selected nuclei. The PECs of $^{178-192}$Hg, $^{185-194}$Pb and $^{196-206}$Po,  as a function of $\beta_2$,  are shown in Figs.~(\ref{pecs_hg}--~\ref{pecs_po}), respectively. In these figures, the binding energy was set to zero at the minimum of each curve. It is noted that the PECs of the  isotopic chains, calculated with D$^3$C, DD-ME2 and DD-PC1 functionals, are quite similar to each other. Fig.~\ref{pecs_hg} shows that the lowest energy configurations of $^{178}$Hg and $^{179}$Hg,  using the three interactions, were obtained  around  $\beta_2=-0.12$.  The RMF model predicted oblate shape for  $^{178}$Hg and $^{179}$Hg. For the case of $^{180-186}$Hg, there were minima on the positive and negative sides of  $\beta_2$ axes resulting in two possible shape configurations (oblate and prolate)  for these nuclei. For $^{187-192}$Hg, the RMF calculation predicted oblate shape configurations for the ground states of these nuclei. The PECs, displayed in Fig.~\ref{pecs_pb} and Fig.~\ref{pecs_po}, show minima at negative values of $\beta_2$ for $^{185-194}$Pb and $^{196-206}$Po, respectively, implying oblate shape configurations for these nuclei. Table~\ref{beta2_Hg} and Table~\ref{beta2_Pb} show the comparison of measured $\beta_2$ values of selected Hg and Pb isotopes, respectively, with the FRDM and RMF calculations. {The measured deformation values were taken from the experimental results of B(E2) values for the known first 2$^+$ states in even–even nuclei~\cite{Pri16}}.  From Table~\ref{beta2_Hg} it is noted that the magnitudes of $\beta_2$  for $^{180}$Hg and $^{182}$Hg using the three different functionals are in good agreement with the experimental data. The prediction of DD-ME2 interaction for $^{188}$Hg is close to measured value. For $^{184}$Hg and $^{186}$Hg, the RMF model predicted bigger magnitude of deformation values using the three functionals. Table~\ref{beta2_Pb} shows a similar situation for  $^{186}$Pb and $^{188}$Pb. The FRDM predictions are almost fifteen times smaller than those of experimental results. Table~\ref{beta2_Po} shows the measured deformation parameters of the ground
state of Po isotopes, extracted from the charge radii (second column) and
sum of squared matrix elements (third column) taken from Ref.~\cite{Kes15}. Calculated values are shown in Columns (4--7). It is noted that the  RMF calculated deformation parameters are in better agreement with the measured data as compared to FRDM values.

Figs.~(\ref{fig:179HgBGT}~-~\ref{fig:197PoBGT}) show the calculated branching ratios and partial half-lives in the left column for the  four selected deformation values. The branching ratio (I) was calculated using
\begin{eqnarray}
	I = \frac{T_{1/2}}{t_{(1/2)}} \times 100 ~(\%).
\end{eqnarray}
The branching ratio increases with decreasing partial half-lives. A bigger  ratio translates to bigger GT strength.  
The right panels of Figs.~(\ref{fig:179HgBGT}~-~\ref{fig:197PoBGT})  show GT strength distributions as a function of daughter excitation energy for odd-A nuclei $^{179}$Hg, $^{185}$Pb and $^{197}$Po, respectively. The distributions are shown within the Q-window. 
The GT strength distributions using the selected deformations (D$^3$C, FRDM, DD-ME2 and DD-PC1) are shown separately in the figures. It is noted that the computed GT strength is well fragmented over the selected energy range for all deformation values. The distributions change appreciably as one moves from FRDM to RMF calculated deformation values. There is a minimal change in the RMF calculated strength distributions. This is due to the reason that the RMF computed $\beta_2$ values are quite close to each other. In Fig.~\ref{fig:185PbBGT}, we notice a small GT strength in the top panel. The FRDM model calculates essentially a spherical shape for $^{185}$Pb and a small strength of 0.00139 within the Q-window.
The change in distributions led to  shifting of the centroid and alteration in total strength of the calculated GT distributions which we discuss next.

Figs.~(\ref{fig:HgHL}~-~\ref{fig:PoHL}) show four panels. Panel (a):  Theoretical estimates of the ground state $\beta_2$ values, panel (b): calculated and measured $\beta$-decay half-lives, panel (c): calculated total GT strength values, and panel (d):  computed centroids of the calculated GT strength distributions.  As stated earlier, the deformed Skyrme HF+BCS+QRPA approach presented only estimated values of deformation parameters. The same was extracted from their data presented in Ref.~\cite{Mor06} and shown  as SLy4 in Figs.~(\ref{fig:HgHL}~-~\ref{fig:PoHL}).   The FRDM calculated  $\beta_2$ values were taken from Ref.~\cite{moller16}. Panel (a) in Figs.~(\ref{fig:HgHL}~-~\ref{fig:PoHL}) show that the FRDM calculated  $\beta_2$ values are comparatively smaller than those computed by other models. Minor variations in total GT strength and centroid values are noted between  different model calculations (see panels (c) and (d)). However, for few cases the differences are substantial. Fig.~\ref{fig:HgHL}(c) shows up to two orders of magnitude difference between the total GT strength calculated by FRDM and other models, for the case of $^{190}$Hg. Similarly more than a factor 2 (5) difference is noted between the FRDM calculated total GT strength of $^{185}$Pb ($^{198}$Po) and other models in Fig.~\ref{fig:PbHL}(c) (Fig.~\ref{fig:PoHL}(c)).  A factor 2 difference in computed centroid values is noted in Fig.~\ref{fig:HgHL}(d) (Fig.~\ref{fig:PbHL}(d)) for the case of $^{179}$Hg ($^{189}$Pb) between FRDM and RMF calculations. We attribute these differences to the changing values of $\beta_2$.

 Figs.~(\ref{fig:HgHL}~-~\ref{fig:PoHL})(b) shows the measured and pn-QRPA calculated half-lives for Hg, Pb and Po isotopes, respectively. Experimental data was taken from Ref.~\cite{Aud21}. All theoretical calculations were performed using the pn-QRPA model with a separable and schematic interaction. All model parameters were kept constant and only the $\beta_2$ values, calculated using the interactions shown in legend, were changed. The overall comparison of calculated and measured data is satisfactory (within a factor 2) with a few exceptions. For the case of $^{185}$Pb, all predicted $\beta$-decay half-life values are around a factor 70 bigger than the measured value. This large difference is attributed to the parameterization of $\chi$ and $\kappa$ values \cite{Hom96}. There are always a few odd cases where deviation in the pn-QRPA model predicted half-lives exceed factor 10 (see Table III of Ref.~\cite{Hom96}). The half-life values vary considerably with changing $\beta_2$ values. Of special mention is the case of  $^{206}$Po. The $\beta_2$ values calculated using the D$^3$C, DD-PC1, DD-ME2 and FRDM are -0.04651, -0.04562, -0.04521 and -0.01481, respectively. The ratios of calculated to measured half-lives are 10.10, 0.89, 0.19 and 1.01, respectively. The half-life value calculated using the D$^3$C interaction is a factor 55 bigger than the value calculated using the DD-PC1 interaction. The reason for this big change may be traced back to the values of total strength and centroid  of the resulting GT  distributions. The D$^3$C interaction resulted in a total strength of 0.0006 and centroid of 1.7 MeV whereas the corresponding values obtained using the DD-PC1 functional were 0.0049 and 1.39 MeV, respectively. A bigger value of total strength and smaller magnitude of GT centroid translate into shorter half-lives. 
 There are also other cases (e.g. $^{179,181,190,191,192}$Hg, $^{189}$Pb, $^{196,198,200,202,203}$Po) where the half-lives changes by more than an order of magnitude as the deformation parameter changes.

Table~\ref{tab:addlabel} displays the deviation of the calculated half-lives from the measured ones and was determined using the relation
\begin{equation}\label{err}
	err({T_{1/2}})= \frac{(T_{1/2}^{Exp} - T_{1/2}^{pnQRPA})}{T_{1/2}^{Exp}}.
\end{equation}
The standard deviations of the err({T$_{1/2}$) values were computed using
	\begin{equation}
		\sigma_{err}= \sqrt{\frac{\sum{[err({T_{1/2})]^2}}}{n}},
	\end{equation}
	and are shown in Table~\ref{tab:addlabel}. It is noted that the SLy4 interaction resulted in the smallest standard deviation for all nuclear configurations. However it should be noted  that the self-consistent deformed
	Skyrme Hartree-Fock calculation was possible only for even-even nuclei. Amongst the RMF interactions, the DD-ME2  gave best results. The FRDM comparison was exceptional for Po isotopes.

\section{Conclusions}
In this work, the ground state deformation parameters of $^{178-192}$Hg, $^{185-194}$Pb and $^{196-206}$Po were investigated using the quadrupole moment constrained RMF calculation. The  D$^3$C, DD-ME2 and DD-PC1 functionals produced similar PECs for the selected nuclei. The results indicate that $^{185-194}$Pb and $^{196-206}$Po isotopes have oblate shape configuration. Regarding Hg nuclei, oblate shape was predicted for $^{178-192}$Hg while two possible shape configurations (oblate and prolate) were predicted for  $^{180-186}$Hg. 

{The following conclusions may be drawn:}

$\bullet$ Calculated half-lives using deformation values from self-consistent deformed
Skyrme Hartree-Fock with SLy4 interaction are in excellent agreement with the measured data. The FRDM deformations also resulted in very good overall agreement with measured half-lives.

$\bullet$ Calculated half-lives for even-even Hg, Pb and odd-A Po isotopes using deformation values from RMF model are in good agreement with the measured data. Amongst the three RMF interactions, the DD-ME2 deformations resulted in best comparison with the measured data.

$\bullet$ The GT strength distributions alter significantly with change in deformation values. This finding validates the conclusion given in Ref.~\cite{Mor06}. 

$\bullet$ Contrary to the findings of Ref.~\cite{Mor06}, it was concluded that \textit{both} the total GT strengths (within the Q-window) and half-lives vary considerably (up to an order of magnitude) with deformation values. They might be considered as good observables to study the deformation effects in nuclei.

\section*{Acknowledgements}
J.-U. Nabi and M. Riaz would like to acknowledge the support of the Higher Education Commission Pakistan through 
Project \# 20-15394/NRPU/R\&D/HEC/2021.


\clearpage
\begin{figure}[h!]
	\hskip-20pt
	\includegraphics[width=1.\textwidth,height=6in]{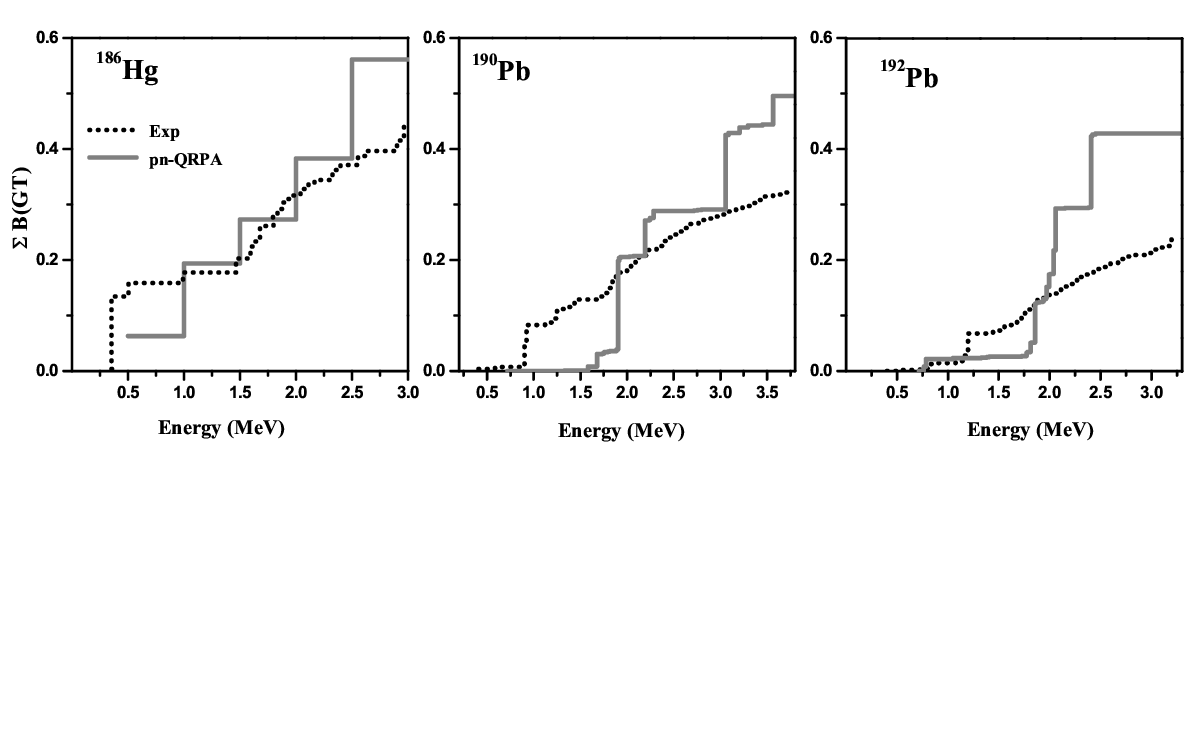}
	\caption{Comparison of the pn-QRPA calculated GT strength distributions of $^{186}$Hg and $^{190,192}$Pb with experimental data~\cite{Alg21,Agu15}.}
	\label{Rb}
\end{figure}
\clearpage
\begin{figure}[h!]
	\includegraphics[width=0.8\textwidth,height=9in]{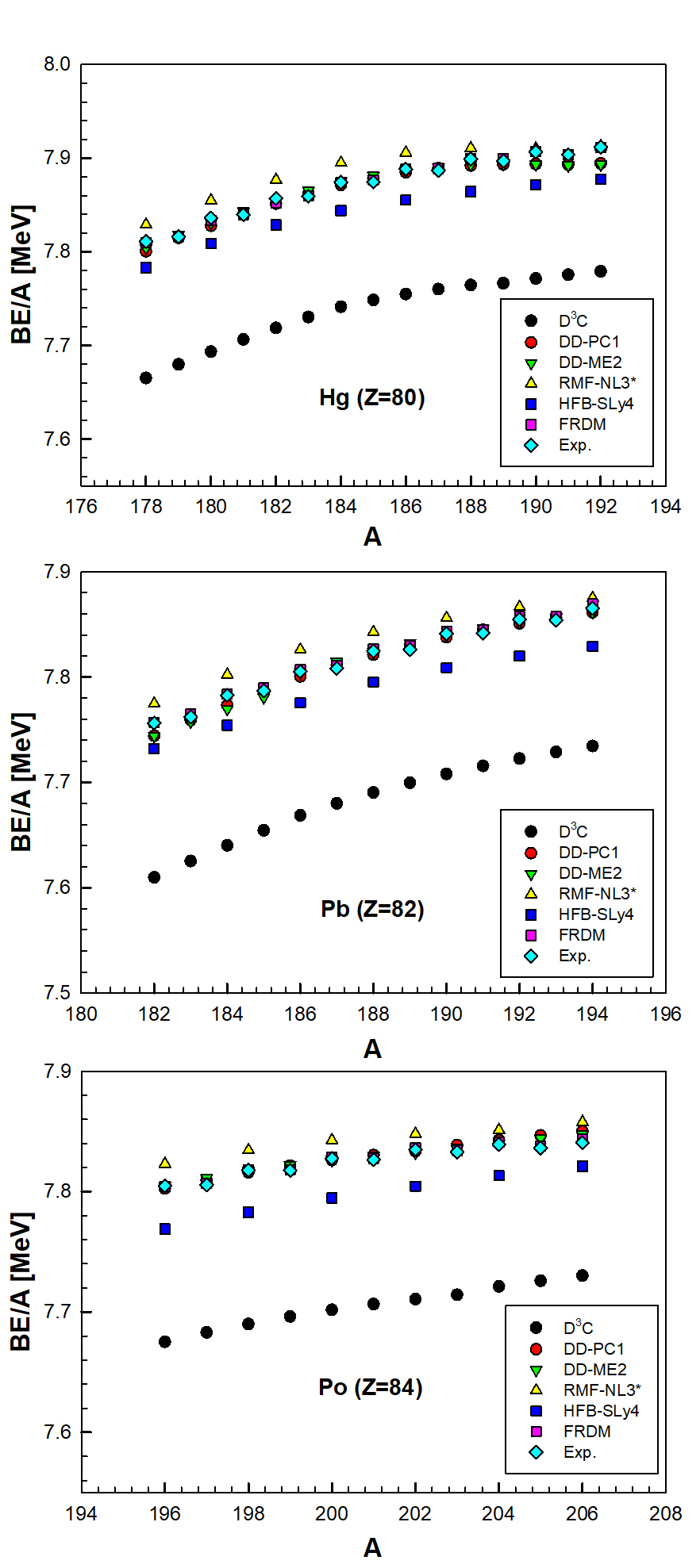}
	\caption{Calculated binding energies per nucleon for Hg, Pb and Po isotopes using the density-dependent D$^3$C, DD-PC1 and DD-ME2 functionals. The results of RMF model with non-linear NL3* parameter set~\cite{bayram13a}, HFB theory with SLy4 interaction~\cite{sto03}, FRDM~\cite{moller16} and experimental data~\cite{Aud21} are also shown.} 
	\label{bea_all}
\end{figure}

\clearpage
\begin{figure}[h!]
	\includegraphics[width=1\textwidth,height=9in]{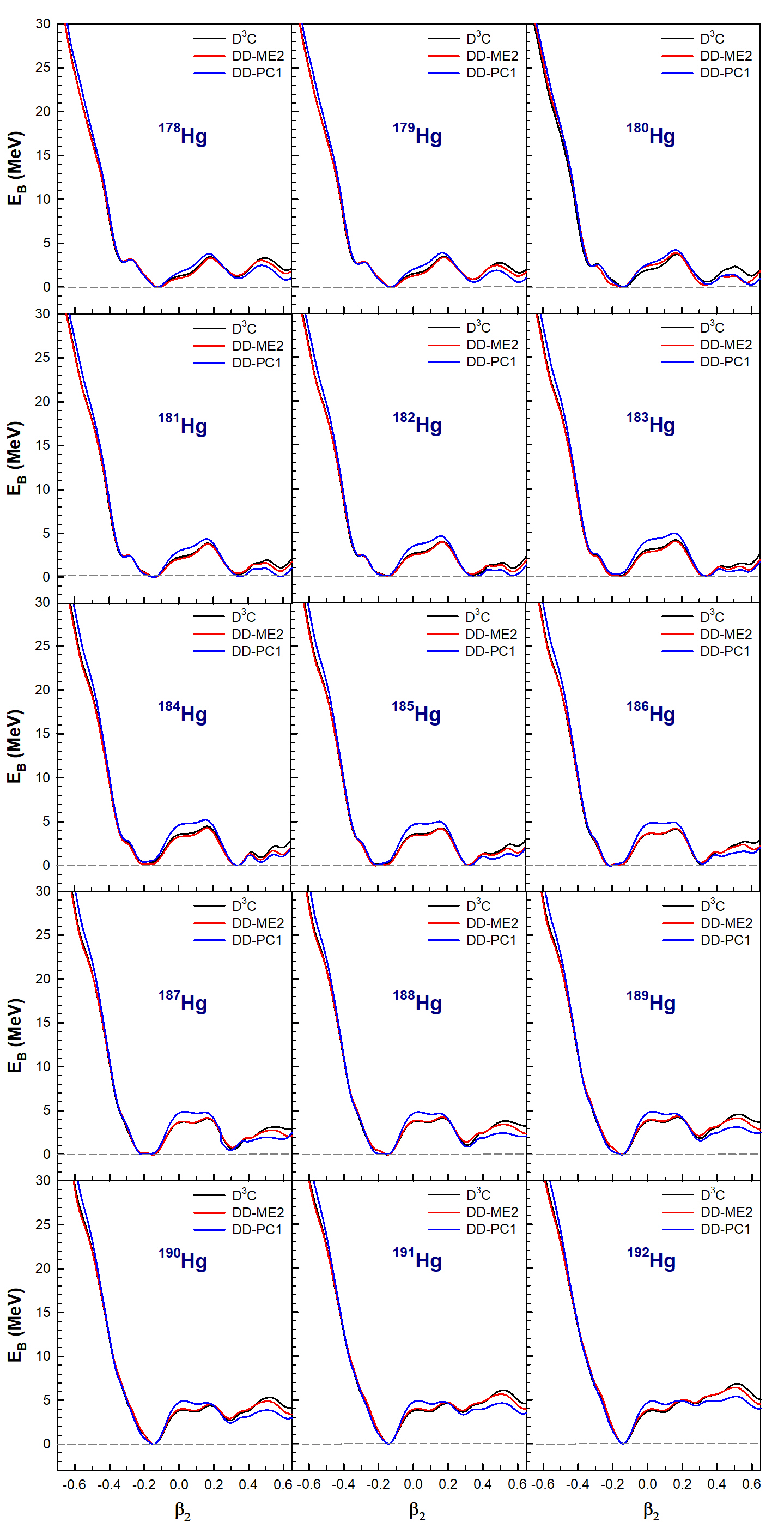}
	\caption{Calculated PECs for $^{178-192}$Hg  using density-dependent D$^3$C, DD-ME2 and DD-PC1 functionals.}
	\label{pecs_hg}
\end{figure}
\clearpage
\begin{figure}[h!]
	\includegraphics[width=1\textwidth,height=9in]{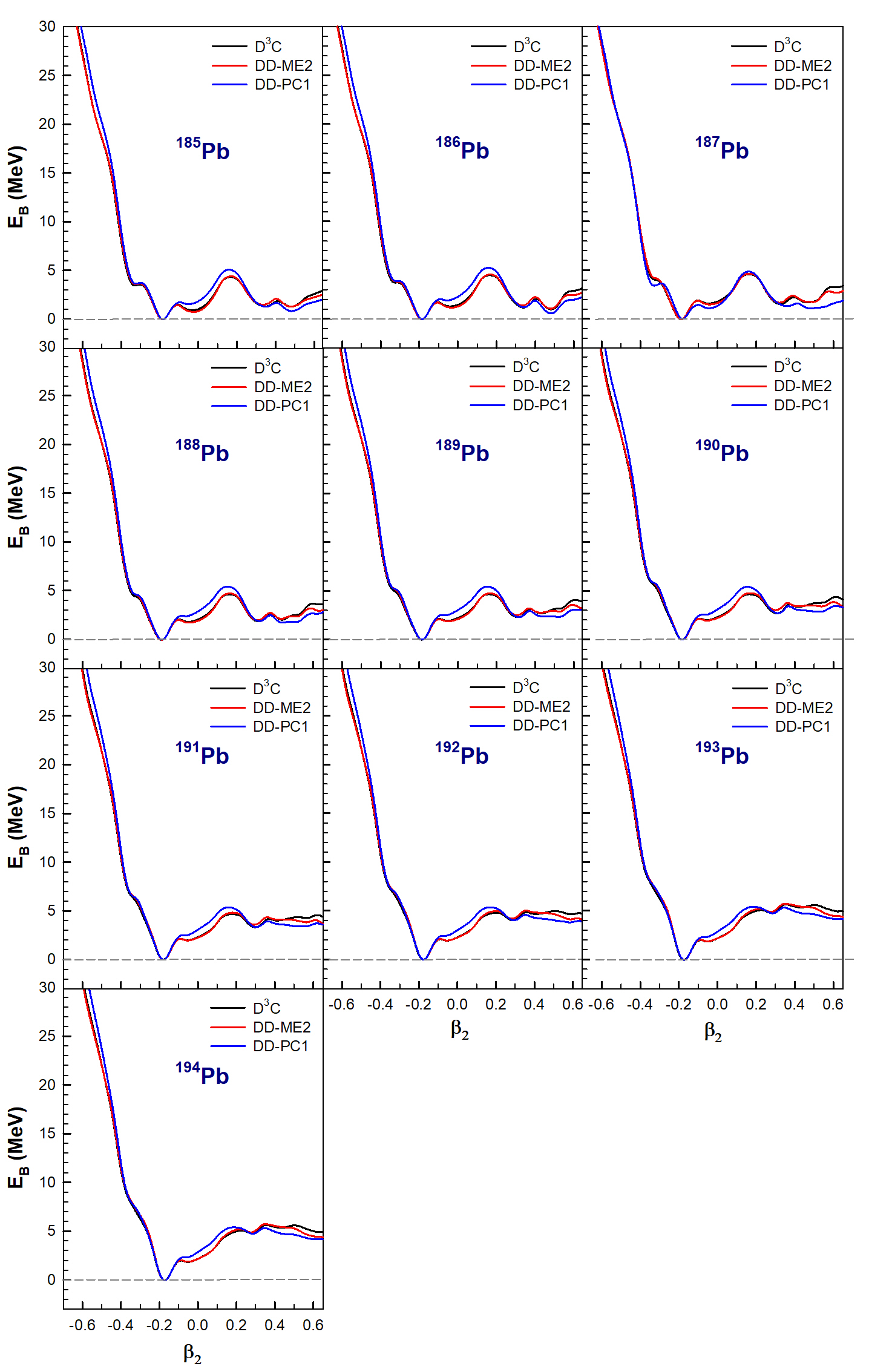}
	\caption{Same as Fig.~\ref{pecs_hg} but for Pb isotopes.}
	\label{pecs_pb}
\end{figure}
\clearpage
\begin{figure}[h!]
	\includegraphics[width=1\textwidth,height=9in]{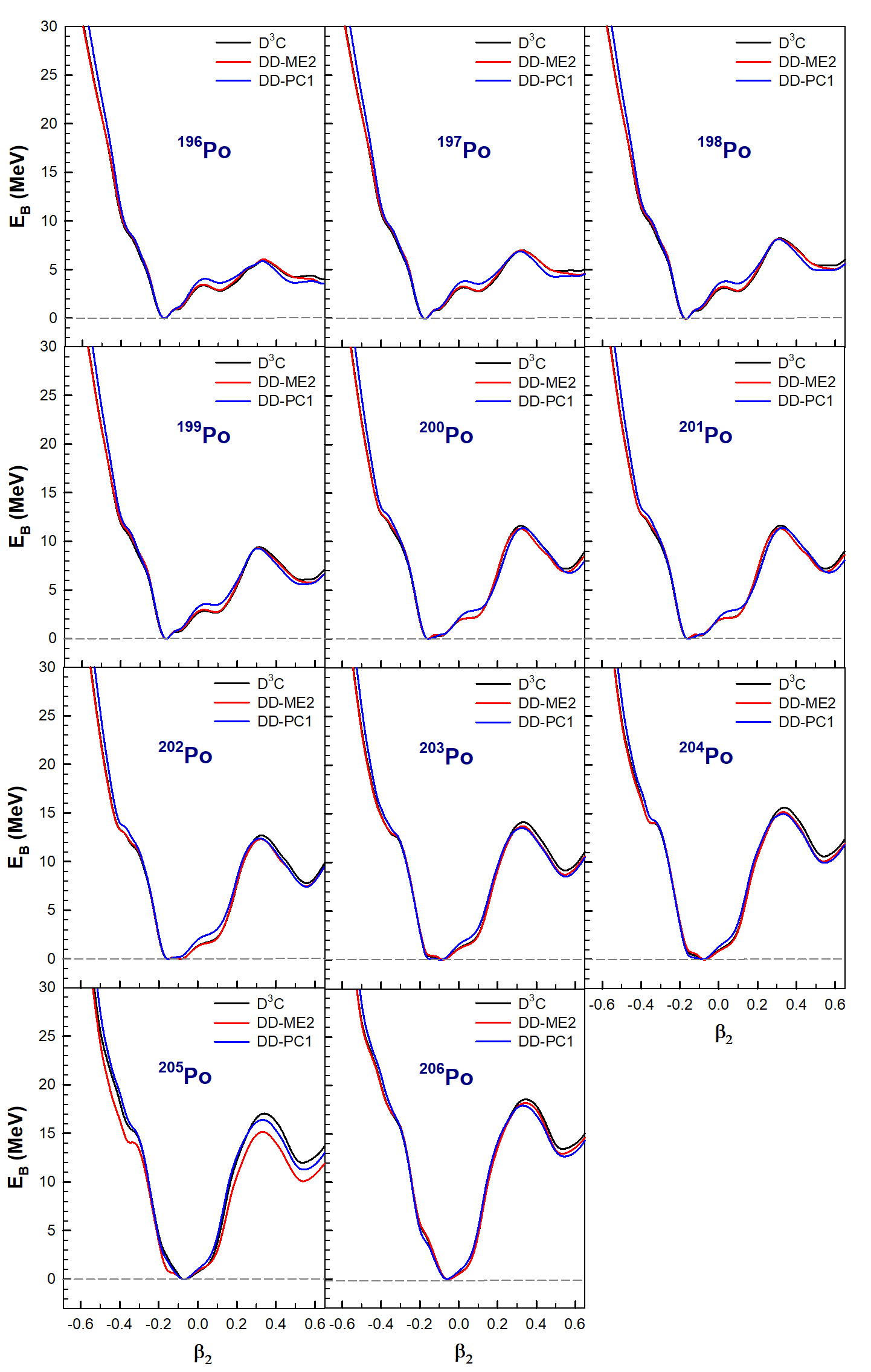}
	\caption{Same as Fig.~\ref{pecs_hg} but for Po isotopes.}
	\label{pecs_po}
\end{figure}
\clearpage

\begin{figure}[h!]
	\hskip-100pt
	\includegraphics[width=1.3\textwidth,height=9in]{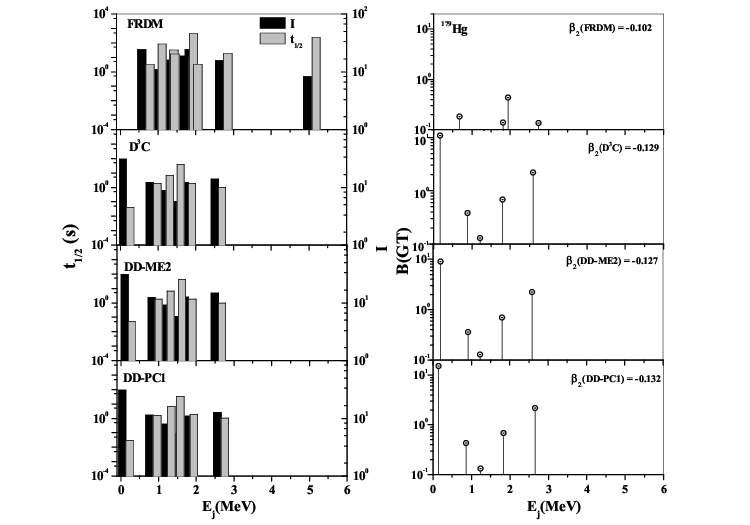}
	\caption{Left column: The pn-QRPA calculated partial  half-lives and branching ratios of  $^{179}$Hg for the four selected deformation values. Right column: The corresponding  GT$_{+}$ strength distributions for the ground state of $^{179}$Hg.}
	\label{fig:179HgBGT}
\end{figure}
\clearpage
\begin{figure}[h!]
	\hskip-70pt
	\includegraphics[width=1.65\textwidth,height=9in]{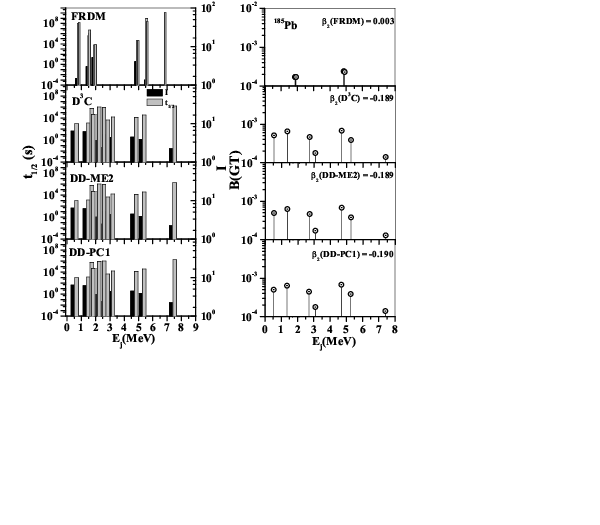}
	\caption{Same as Fig.~\ref{fig:179HgBGT} but for $^{185}$Pb.}
	\label{fig:185PbBGT}
\end{figure}
\clearpage
\begin{figure}[h!]
	\hskip-45pt
	\includegraphics[width=1.55\textwidth,height=9in]{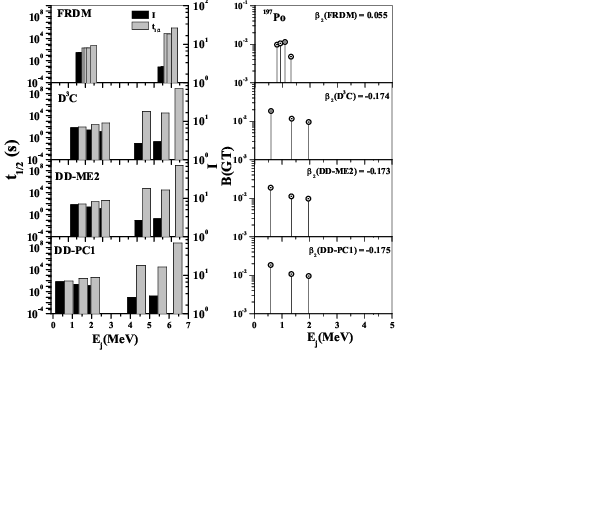}
	\caption{Same as Fig.~\ref{fig:179HgBGT} but for $^{197}$Po.}
	\label{fig:197PoBGT}
\end{figure}

\clearpage

\begin{figure}[h!]
	\hskip-90pt
	\includegraphics[width=1.75\textwidth,height=9.4in]{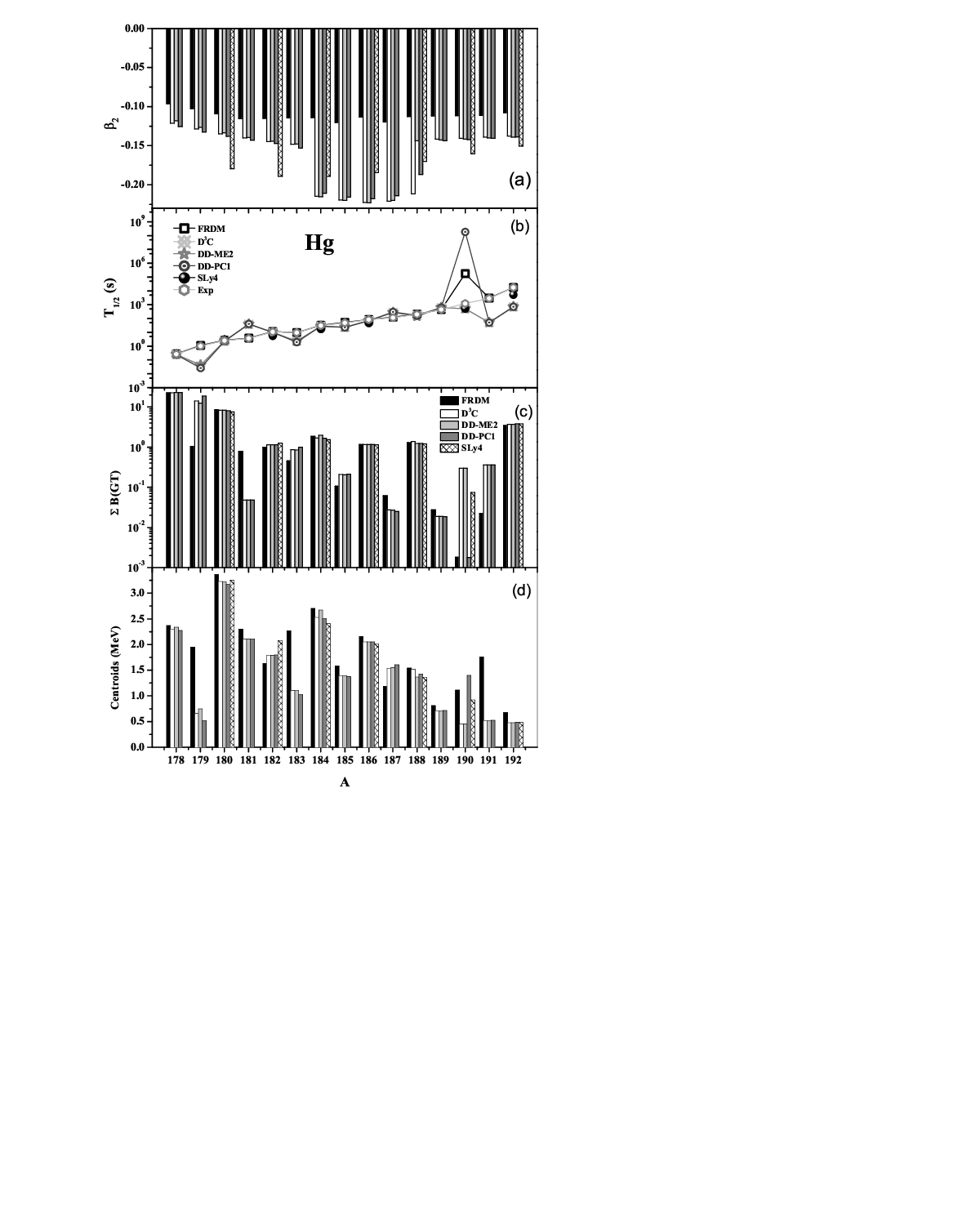}
	\caption{Calculated $\beta$-decay properties of Hg isotopes using five different interactions shown in legends. (a) $\beta_2$, (b)  $T_{1/2}$ , (c) total GT$_+$ strength  and (d) centroids of GT distributions. Measured half-lives were taken from Ref.~\cite{Aud21}.}
	\label{fig:HgHL}
\end{figure}
\clearpage

\begin{figure}[h!]
	\hskip-65pt
	\includegraphics[width=2.3\textwidth,height=9.5in]{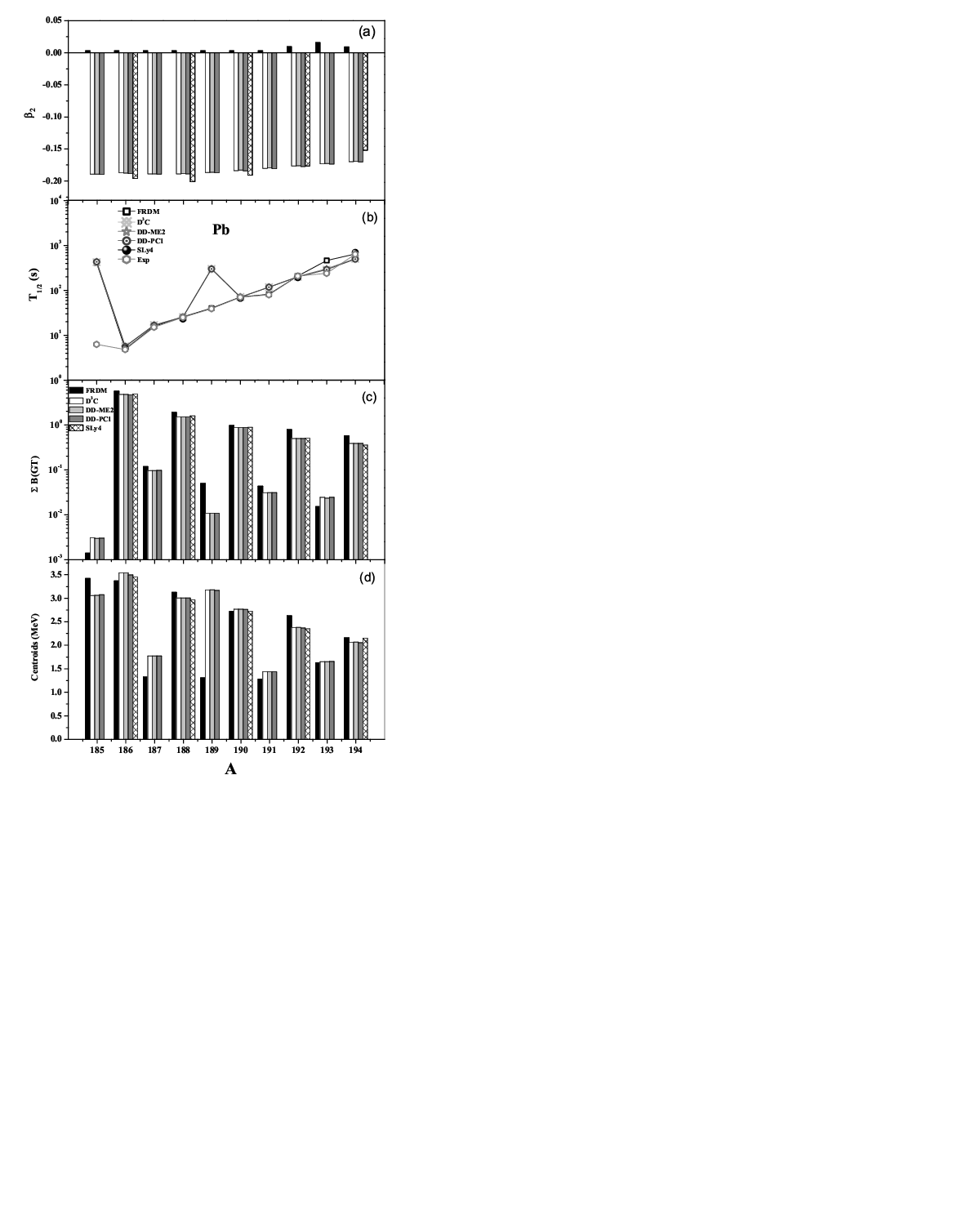}
	\caption{Same as Fig.~\ref{fig:HgHL} but for Pb isotopes.}
	\label{fig:PbHL}
\end{figure}

\clearpage

\begin{figure}[h!]
\hskip-85pt
\includegraphics[width=1.0\textwidth,height=8.5in]{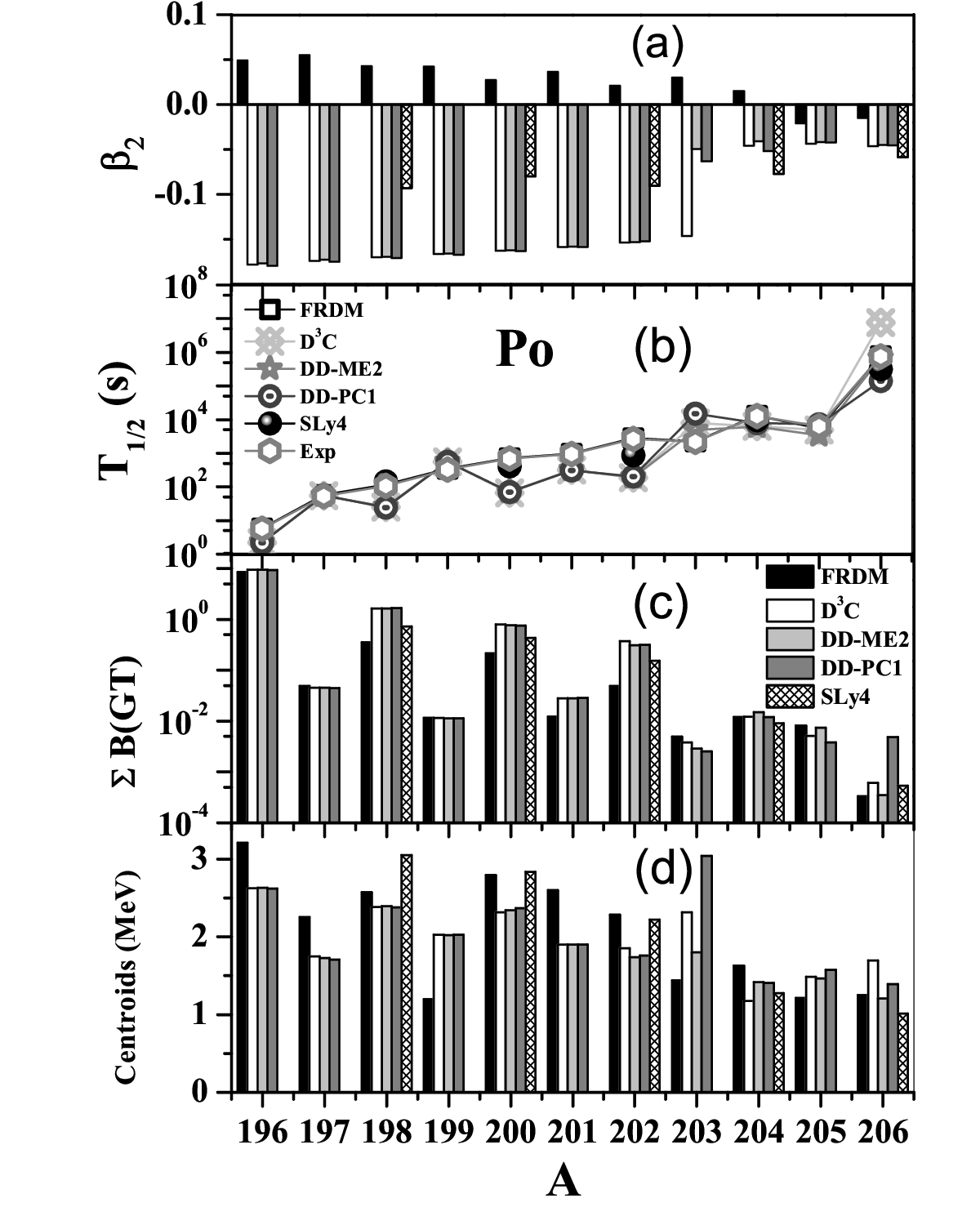}
\caption{Same as Fig.~\ref{fig:HgHL} but for Po isotopes.}
\label{fig:PoHL}
\end{figure}

\clearpage

\begin{table}[htbp]
	\centering
	\caption{Comparison of measured and calculated $\beta_2$ values of Hg isotopes. Numbers in parenthesis show uncertainties in the extraction process.}
	\begin{tabular}{c|ccccc}
	\hline
	&\multicolumn{5}{|c}{\textbf{$\beta_2$}} \\
	\hline
	&&&&&\\
	A     & Exp~\cite{Pri16} &  FRDM~\cite{moller16}  & D$^3$C  & DD-ME2   & DD-PC1 \\
	\hline
	180   & 0.137(99) & -0.109 & -0.135 & -0.134 & -0.138 \\
	182   & 0.147(38) & -0.115 & -0.145 & -0.144 & -0.147 \\
	184   & 0.143(03) & -0.114 & -0.215 & -0.215 & -0.211 \\
	186   & 0.135(10) & -0.113 & -0.223 & -0.223 & -0.218 \\
	188   & 0.145(13) & -0.112 & -0.212 & -0.143 & -0.187 \\
	\hline
\end{tabular}

	\label{beta2_Hg}
\end{table}%

\begin{table}[htbp]
	\centering
	\caption{Comparison of measured and calculated $\beta_2$ values of Pb isotopes. Numbers in parenthesis show uncertainties in the extraction process.}
	\begin{tabular}{c|ccccc}
	\hline
	&\multicolumn{5}{|c}{\textbf{$\beta_2$}} \\
	\hline
	&&&&&\\
	A     & Exp~\cite{Pri16} &  FRDM~\cite{moller16}  & D$^3$C  & DD-ME2   & DD-PC1 \\
	\hline
	186   & 0.048(66) & 0.003 & -0.187 & -0.188 & -0.188 \\
	188   & 0.055(91) & 0.003 & -0.189 & -0.188 & -0.189 \\
	\hline
\end{tabular}

	\label{beta2_Pb}
\end{table}%

\begin{table}[htbp]
	\centering
	\caption{Comparison of measured and calculated $\beta_2$ values of Po isotopes. Numbers in parenthesis show uncertainties in the extraction process.}
	\begin{tabular}{c|cccccc}
	\hline
	&\multicolumn{6}{|c}{\textbf{$\beta_2$}} \\
	\hline		&&&&&&\\
	A     & $\delta\left\langle r^2\right\rangle $ ~\cite{Kes15} & B(E2)~\cite{Kes15} & D$^3$C   & DD-ME2   & DD-PC1   & FRDM~\cite{moller16}  \\
	\hline
	196   & 0.159 & 0.129(05) & -0.178 & -0.177 & -0.180 & 0.049 \\
	197   & 0.130 &   -   & -0.174 & -0.173 & -0.175 & 0.055 \\
	198   & 0.119 & 0.116(14) & -0.170 & -0.169 & -0.171 & 0.043 \\
	199   & 0.090 &   -   & -0.167 & -0.166 & -0.167 & 0.042 \\
	200   & 0.100 & 0.104(03) & -0.163 & -0.162 & -0.163 & 0.027 \\
	201   & 0.100 &   -   & -0.159 & -0.158 & -0.159 & 0.036 \\
	202   & 0.090 & 0.106(14) & -0.154 & -0.153 & -0.152 & 0.021 \\
	
	\hline
\end{tabular}

	\label{beta2_Po}
\end{table}%

\begin{table*}[htbp]
	\centering
	\caption{Standard deviations of the err(T$_{1/2}$) [Eq.~(43)]  using different interactions. }
	\begin{tabular}{rr|rrrrr}
		\hline
		&       & \multicolumn{5}{c}{$\sigma_{err}$} \\
		\hline
		Nuclei & Shape & FRDM  & D$^3$C   & DD-ME2   & DD-PC1   & SLy4 \\
		\hline
		Hg    & Oblate & \multicolumn{1}{c}{2.863} & \multicolumn{1}{c}{2.663} & \multicolumn{1}{c}{2.652} & \multicolumn{1}{c}{2.694} & \multicolumn{1}{c}{1.762} \\
		& Prolate &\multicolumn{1}{c}{-}&1.151&1.151&0.811&0.711\\
		Pb    & Oblate & \multicolumn{1}{c}{20.197} & \multicolumn{1}{c}{21.282} & \multicolumn{1}{c}{21.680} & \multicolumn{1}{c}{21.452} & \multicolumn{1}{c}{0.082} \\
		& Prolate & \multicolumn{1}{c}{-} & \multicolumn{1}{c}{-} & \multicolumn{1}{c}{-} & \multicolumn{1}{c}{-} & \multicolumn{1}{c}{0.022} \\
		& Spherical  & \multicolumn{1}{c}{-} & \multicolumn{1}{c}{-} & \multicolumn{1}{c}{-} & \multicolumn{1}{c}{-} & \multicolumn{1}{c}{0.021} \\
		Po    & Oblate & \multicolumn{1}{c}{0.062} & \multicolumn{1}{c}{2.913} & \multicolumn{1}{c}{0.714} & \multicolumn{1}{c}{1.890} & \multicolumn{1}{c}{0.483} \\
		& Prolate & \multicolumn{1}{c}{-} & \multicolumn{1}{c}{-} & \multicolumn{1}{c}{-} & \multicolumn{1}{c}{-} & \multicolumn{1}{c}{0.543} \\
		\hline
	\end{tabular}%
	\label{tab:addlabel}%
\end{table*}%

    \end{document}